\documentclass[aps,prd,superscriptaddress,twocolumn,floatfix,longbibliography]{revtex4-2}
\usepackage{graphicx}
\usepackage{amsmath}
\usepackage{float}
\usepackage{appendix}
\usepackage{xcolor}
\usepackage{hyperref}% add hypertext capabilities
\usepackage{siunitx}
\hypersetup{
    colorlinks,
    linkcolor={red!50!black},
    citecolor={blue!50!black},
    urlcolor={blue!80!black}
}
\usepackage{lineno}
\newcommand{\degc}{$^\circ\text{C}$}
\newcommand{\ybe}{$^{88}$Y-Be}
\newcommand{\bibe}{$^{207}$Bi-Be}
\newcommand{\cf}{{$^{252}$Cf}}
\newcommand{\kevnr}{keV$_{nr}$}
\newcommand{\co}{$^{57}$Co}
\newcommand{\bi}{$^{207}$Bi}
\newcommand{\y}{$^{88}$Y}
\newcommand{\qseitz}{$Q_{\mathrm{Seitz}}$}
\newcommand{\pom}{~$\pm$~}

\begin{document}

\title{Low-threshold response of a scintillating xenon bubble chamber to nuclear and electronic recoils
}

\date{\today}

% The SBC Collaboration Author List

%alphabetical affiation list?
\affiliation{Department of Physics, University of Alberta, Edmonton, T6G 2E1, Canada}
\affiliation{Department of Physics and Astronomy, University of California Riverside, Riverside, California 92507, USA}
\affiliation{Department of Physics, University of California Santa Barbara, Santa Barbara, California, 93106, USA}
\affiliation{Department of Physics, Drexel University, Philadelphia, Pennsylvania 19104, USA}
\affiliation{Fermi National Accelerator Laboratory, Batavia, Illinois 60510, USA}
\affiliation{Department of Physics, Indiana University South Bend, South Bend, Indiana 46634, USA}
%\affiliation{Department of Chemical Engineering and Materials Science, University of Minnesota, Minneapolis 55455, USA}
\affiliation{D\'{e}partement de Physique, Universit\'{e} de Montr\'{e}al, Montr\'{e}al, H3T 1J4, Canada}
\affiliation{Instituto de F\'{\i}sica, Universidad Nacional Aut\'onoma de M\'exico, A.P. 20-364, Ciudad de M\'exico 01000, M\'exico}
\affiliation{Northeastern Illinois University, Chicago, Illinois 60625, USA}
\affiliation{Department of Physics and Astronomy, Northwestern University, Evanston, Illinois 60208, USA}
\affiliation{Department of Physics, Queen's University, Kingston, K7L 3N6, Canada}
\affiliation{SNOLAB, Lively, Ontario, P3Y 1N2, Canada}
\affiliation{TRIUMF,  Vancouver,  BC  V6T  2A3,  Canada}

\author{E.~Alfonso-Pita}
\affiliation{Instituto de F\'{\i}sica, Universidad Nacional Aut\'onoma de M\'exico, A.P. 20-364, Ciudad de M\'exico 01000, M\'exico}

%\author{M.~Baker}
%\affiliation{Department of Physics, University of Alberta, Edmonton, T6G 2E1, Canada}

\author{E.~Behnke}
\affiliation{Department of Physics, Indiana University South Bend, South Bend, Indiana 46634, USA}

\author{M.~Bressler}
\email[Contact author: ]{mbressler@umass.edu}
\altaffiliation[Present address: ]{University of Massachusetts, Amherst, 01002, USA.}
\affiliation{Department of Physics, Drexel University, Philadelphia, Pennsylvania 19104, USA}

\author{B.~Broerman}
\affiliation{Department of Physics, Queen's University, Kingston, K7L 3N6, Canada}

\author{K.~Clark}
\affiliation{Department of Physics, Queen's University, Kingston, K7L 3N6, Canada}

\author{R.~Coppejans}
\affiliation{Department of Physics and Astronomy, Northwestern University, Evanston, Illinois 60208, USA}

\author{J.~Corbett}
\affiliation{Department of Physics, Queen's University, Kingston, K7L 3N6, Canada}

\author{M.~Crisler}
\affiliation{Fermi National Accelerator Laboratory, Batavia, Illinois 60510, USA}
%\affiliation{Pacific Northwest National Laboratory,\ Richland,\ Washington\ 99354,\ USA}

\author{C.~E.~Dahl}
\affiliation{Department of Physics and Astronomy, Northwestern University, Evanston, Illinois 60208, USA}
\affiliation{Fermi National Accelerator Laboratory, Batavia, Illinois 60510, USA}

\author{K.~Dering}
\affiliation{Department of Physics, Queen's University, Kingston, K7L 3N6, Canada}

\author{A.~de St.~Croix}
\affiliation{Department of Physics, Queen's University, Kingston, K7L 3N6, Canada}

\author{D.~Durnford}
\email[Contact author: ]{ddurnfor@ualberta.ca}
\affiliation{Department of Physics, University of Alberta, Edmonton, T6G 2E1, Canada}

\author{P.~Giampa}
\affiliation{TRIUMF,  Vancouver,  BC  V6T  2A3,  Canada}

\author{J.~Hall}
\affiliation{SNOLAB, Lively, Ontario, P3Y 1N2, Canada}

\author{O.~Harris}
\affiliation{Northeastern Illinois University, Chicago, Illinois 60625, USA}

\author{H.~Hawley-Herrera}
\affiliation{Department of Physics, Queen's University, Kingston, K7L 3N6, Canada}

%\author{Y.~Ko}
%\affiliation{Department of Physics, University of Alberta, Edmonton, T6G 2E1, Canada}

\author{N.~Lamb}
\affiliation{Department of Physics, Drexel University, Philadelphia, Pennsylvania 19104, USA}

\author{M.~Laurin}
\affiliation{D\'{e}partement de Physique, Universit\'{e} de Montr\'{e}al, Montr\'{e}al, H3T 1J4, Canada}

\author{I.~Levine}
\affiliation{Department of Physics, Indiana University South Bend, South Bend, Indiana 46634, USA}

\author{W.~H.~Lippincott}
\affiliation{Department of Physics, University of California Santa Barbara, Santa Barbara, California, 93106, USA}

\author{R.~Neilson}
\affiliation{Department of Physics, Drexel University, Philadelphia, Pennsylvania 19104, USA}

\author{M.-C.~Piro}
\affiliation{Department of Physics, University of Alberta, Edmonton, T6G 2E1, Canada}

%\author{S.~Priya}
%\affiliation{Department of Chemical Engineering and Materials Science, University of Minnesota, Minneapolis 55455, USA}

\author{D.~Pyda}
\affiliation{Department of Physics, Drexel University, Philadelphia, Pennsylvania 19104, USA}

\author{Z.~Sheng}
\affiliation{Department of Physics and Astronomy, Northwestern University, Evanston, Illinois 60208, USA}

\author{G.~Sweeney}
\affiliation{Department of Physics, Queen's University, Kingston, K7L 3N6, Canada}

\author{E.~V\'{a}zquez-J\'{a}uregui}
\affiliation{Instituto de F\'{\i}sica, Universidad Nacional Aut\'onoma de M\'exico, A.P. 20-364, Ciudad de M\'exico 01000, M\'exico}

%\author{D.~Velasco}
%\affiliation{Department of Physics and Astronomy, Northwestern University, Evanston, Illinois 60208, USA}

\author{S.~Westerdale}
\affiliation{Department of Physics and Astronomy, University of California Riverside, Riverside, California 92507, USA}

\author{T.~J.~Whitis}
\affiliation{Department of Physics, University of California Santa Barbara, Santa Barbara, California, 93106, USA}

\author{A.~Wright}
\affiliation{Department of Physics, Queen's University, Kingston, K7L 3N6, Canada}

\author{J.~Zhang}
\affiliation{Department of Physics and Astronomy, Northwestern University, Evanston, Illinois 60208, USA}

\author{R.~Zhang}
\affiliation{Department of Physics, University of California Santa Barbara, Santa Barbara, California, 93106, USA}

\author{A.~Zu\~niga-Reyes}
 \altaffiliation[Present address: ]{University of Toronto, Toronto, M5S 1A7, Canada.}
\affiliation{Instituto de F\'{\i}sica, Universidad Nacional Aut\'onoma de M\'exico, A.P. 20-364, Ciudad de M\'exico 01000, M\'exico}

\collaboration{SBC collaboration}
	\noaffiliation
%	\email{sbc@snolab.ca}

\begin{abstract}
A device filled with pure xenon first demonstrated the ability to operate simultaneously as a bubble chamber and scintillation detector in 2017. Initial results from data taken at thermodynamic thresholds down to $\sim$4~keV showed sensitivity to $\sim$20~keV nuclear recoils with no observable bubble nucleation by $\gamma$-ray interactions. This paper presents results from further operation of the same device at thermodynamic thresholds as low as $0.50$~keV, hardware limited. The bubble chamber has now been shown to have sensitivity to $\sim$1~keV nuclear recoils while remaining insensitive to bubble nucleation by $\gamma$-rays. A data-driven calibration of the chamber's nuclear recoil nucleation response, as a function of nuclear recoil energy and thermodynamic state, is presented. Stringent upper limits are established for the probability of bubble nucleation by $\gamma$-ray-induced Auger cascades, with a limit of $<1.1\times10^{-6}$ set at $0.50$~keV, the lowest thermodynamic threshold explored. 
\end{abstract}

\maketitle

\section{Introduction}
Bubble chambers have been a staple of particle detection since their invention and study by Glaser, Seitz, and others in the 1950s~\cite{glaser,characteristics,Seitz}, and have been historically operated with a wide variety of fluids~\cite{bugg}. The fluid choice is generally governed by balancing the engineering needs of achieving stable pressures and temperatures (fluid-dependent), and the physics needs of searching for interactions with specific targets or desire for a particular fluid density. An early attempt at creating a xenon bubble chamber was initially unsuccessful, failing to produce $\delta$-electron tracks when exposed to $\gamma$-rays, except when the scintillation was quenched by adding ethylene~\cite{glaserxenon}. Xenon bubble chambers with scintillation-quenching additives later proved useful in several experiments~\cite{xebckdecays}. In these experiments, xenon was not chosen for its scintillation properties, but rather for its high density---desirable to increase the rate of rare events and reduce the radiation length~\cite{bugg}.

More recently, fluorocarbon bubble chambers have been deployed by COUPP, PICO, and others to search for nuclear recoils from dark matter interactions~\cite{COUPP4,SIMPLE:2014pun,PICO:2015pux,Behnke:2016lsk,60complete}. These experiments have operated at moderate degrees of superheat to effectively eliminate bubble nucleation from radiation backgrounds, forgoing sensitivity to the lowest energy nuclear recoils (below a few keV). Bubble chamber dark matter searches would benefit from the properties found accidentally in the original unquenched xenon bubble chamber: a high-density liquid that is insensitive to bubble nucleation by $\gamma$-rays even at high degrees of superheat.

It was shown in Ref.~\cite{XeBC} that pure xenon bubble chambers are sensitive to nuclear recoils, as necessary for a dark matter search, despite being insensitive to $\gamma$-rays. Thus, the scintillating noble liquid bubble chamber is a promising technology for exploring dark matter, in particular $\mathcal{O}$(GeV) dark matter where sensitivity to sub-keV nuclear recoils is necessary, as well as for detection of coherent elastic neutrino-nucleus scattering (CE$\nu$NS)~\cite{sbc_cevnns,sbc_cevns2}. This paper describes further experiments with the same xenon bubble chamber (referred to as ``XeBC'') as the one used in Ref.~\cite{XeBC}, modified for operation at higher degrees of superheat (lower nuclear recoil detection thresholds). A variety of $\gamma$-ray and neutron sources are used to characterize the response of superheated liquid xenon to both electronic and nuclear recoils. Preliminary results from the data presented here~\cite{matt_thesis,ddurnford_thesis} have already motivated the construction of two $10\,\mathrm{kg}$ liquid argon scintillating bubble chambers currently being installed at Fermilab and SNOLAB by the SBC Collaboration~\cite{sbc_snowmass,universe9080346}. The calibration strategies and analysis techniques presented in this work are directly relevant to these chambers, and to other future dark matter direct detection and CE$\nu$NS experiments that rely on measurements with neutron and $\gamma$-ray sources to constrain the detector's low-energy response.

\section{Previous Experiments and Motivation}
\label{sec:previous_work}

The PICO Collaboration has recently published extensive studies of electronic recoil (ER) and nuclear recoil (NR) induced bubble nucleation in C$_3$F$_8$ bubble chambers~\cite{ERpaper, DBCinstrumentationpaper, DBCAuger, PICONR}. The well-established ``Seitz'' threshold~\cite{Seitz,ERpaper} (calculated from the pressure, temperature, and thermophysical properties of the target fluid, and denoted here as \qseitz{}) is used both as a measure of the degree of superheat in these studies, and as an $\mathcal{O}(1)$ indication of the expected NR detection threshold. In practice, this quantity is seen to underestimate both observed~\cite{PICONR, COUPPCF3I, alphaPICASSO, dErrico, dErrico2, Tardif} and simulated~\cite{lammps_paper,bubbleMD} thresholds for NR-induced bubble nucleation. The level of disagreement varies with both target fluid and threshold, requiring NR nucleation efficiencies to be determined in each new case through dedicated neutron calibrations. The data-driven approach developed by the PICO Collaboration~\cite{PICONR} consists of a global fit of simulated to observed bubble nucleation rates with data gathered from multiple neutron calibration sources.

The ability of a bubble chamber to reject ER events is paramount to the scientific goals of dark matter and CE$\nu$NS detection. Bubble chambers are intrinsically less sensitive to ER events than NRs, in part due to the lower stopping power of ERs~\cite{COUPP:2008vjl}. 
%It is also well-established that ERs induced by $\gamma$-rays cause bubble nucleation, albeit with low probability, in chambers with fluorocarbon active fluids when they are operated in certain thermodynamic states corresponding to NR thresholds of interest to dark matter searches \cite{COUPP4}.
The recent PICO ER studies have made the most careful characterizations of bubble chamber $\gamma$-ray response to date as dark matter searches seek to operate at ever lower thresholds~\cite{ERpaper}. One key result of the recent studies is the realization that $\gamma$-ray sensitivity is dominated by photoabsorption (and subsequent Auger cascades) in any molecular fluid containing a high-Z atom (e.g., CF$_3$I~\cite{ERpaper}) or high-Z contaminant (e.g., xenon dissolved in C$_3$F$_8$~\cite{DBCAuger}).  %in C$_3$F$_8$ is somewhat decoupled from the NR sensitivity and that heavy element contamination causes a dramatic increase in $\gamma$-ray sensitivity due to photoabsorption leading to Auger cascades.

In 2017, XeBC observed for the first time coincident bubble nucleation and scintillation from NRs~\cite{XeBC} without the bubble nucleation by photoabsorption seen in xenon-contaminated C$_3$F$_8$. This paper presents new results from low-threshold (high-superheat) XeBC data taken in the year following the 2017 publication, as described in Sec.~\ref{S:methods}, with NR and ER analyses in the style of the above PICO analyses presented in Sec.~\ref{S:results}. Section~\ref{sec:thresholds} presents constraints on the NR nucleation efficiency in xenon at Seitz thresholds between 0.9~and 2.1~keV. This analysis takes an approach similar to Ref.~\cite{PICONR}, making minimal assumptions regarding the dependence of nucleation efficiency on either recoil energy or thermodynamic state. Section~\ref{gammas} gives upper limits on XeBC's sensitivity to ER events at Seitz thresholds down to 0.5~keV. No evidence for bubble nucleation by $\gamma$-ray photoabsorption is seen, extending the disparity between xenon and the molecular fluids studied by PICO.%, the xenon ER response is further investigated at higher degrees of superheat, corresponding to lower NR thresholds and thus sensitivity to lower mass dark matter.

\section{Experimental Methods}
\label{S:methods}
\subsection{XeBC}

\begin{figure*}[t]
    \centering
    \includegraphics[width=\textwidth, trim={0 0 0 0cm},clip]{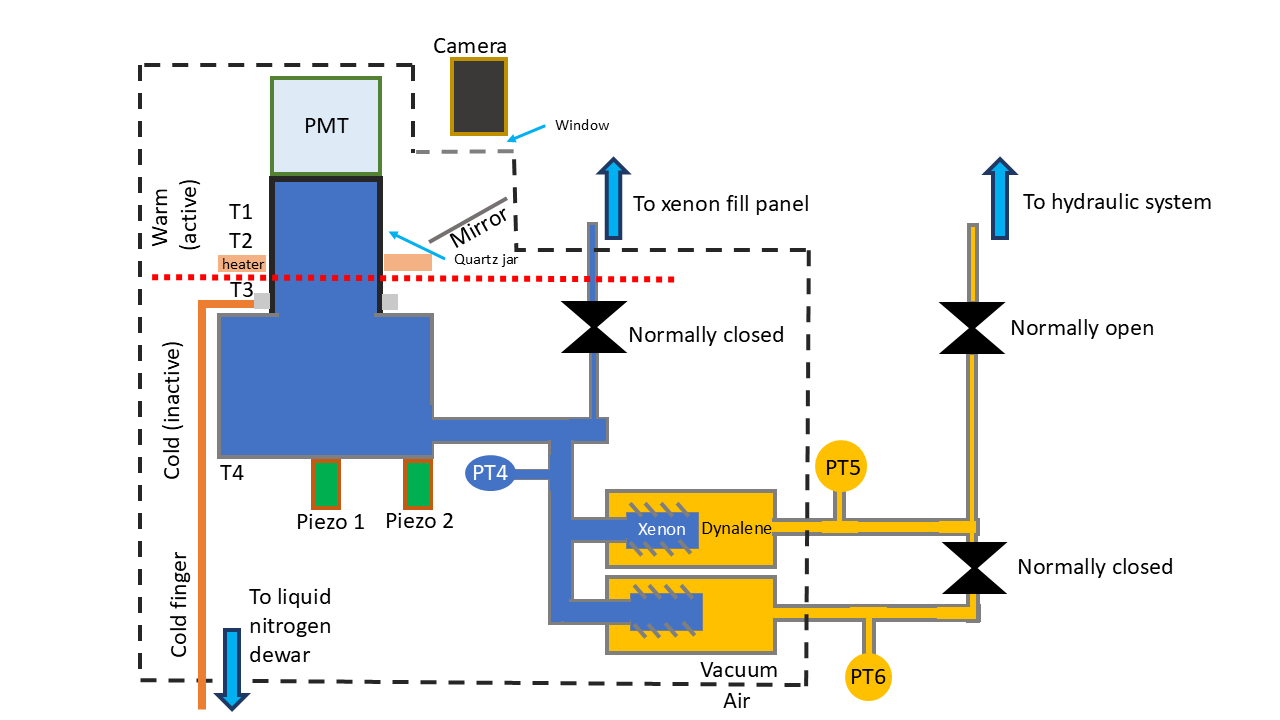}
    \caption{A basic schematic of the xenon bubble chamber's instrumentation and pressure/temperature control scheme. The dashed black line indicates the vacuum cryostat boundary; everything inside it is kept cold by conduction from the cold finger. The red dashed line shows the approximate location of the boundary between cold (inactive) and warm (active) xenon. PT4, PT5, and PT6 are pressure transducers, which provide feedback signals to the hydraulic pressure control system. T1, T2, T3, and T4 are resistive temperature devices (RTDs). %T2 provides feedback to the resistive heater controlling the temperature of the superheated target.
    }
    \label{fig:xebcschematic}
\end{figure*}

The experiments presented here are performed with XeBC, described in Ref.~\cite{XeBC}, with upgrades to enable lower threshold operations. A schematic of the setup is shown in Fig.~\ref{fig:xebcschematic}. The XeBC active target consists of 20~g of superheated liquid xenon, contained in a fused quartz vial within a vacuum cryostat. Bubbles are observed through a glass view port by a video camera outside the cryostat, with mirrors employed to provide a stereo view of the target. A pair of piezoelectric acoustic transducers record the shock waves generated by expanding bubbles. Scintillation photons are counted by a single solar-blind R6834 Hamamatsu photomultiplier tube (PMT) above the target, inside the cryostat and in direct contact with the quartz vial. The superheated xenon is separated from colder stable liquid xenon in the plumbing region below by a steep (60~K over 8~cm) temperature gradient. %A hydraulic system cycles the chamber between a  compressed high-pressure stable liquid state and an expanded low-pressure superheated state. 

For low-threshold operations, an inner quartz vial present in the prior publication was removed from the bubble chamber and replaced with a blank flange at the bottom of the xenon volume. The inner vial served only to displace xenon below the active region, leaving a thin annular region between vials. This region was susceptible to hot spots, and removal of the inner jar was necessary for stability at high degrees of superheat. The outer quartz vial outer diameter was also reduced from 3.0~cm to 2.5~cm (keeping a 3-mm wall) to withstand the higher compression pressures needed at high superheat. These changes increased the total mass of xenon but reduced the volume in the active superheated region. The piezoelectric transducer previously placed inside the inner vial was moved to the lower flange in the cold region, directly below the active liquid target volume, with a second transducer added offset from the chamber axis. After these modifications, XeBC achieved target temperatures up to $-38$\degc{} (limited by the maximum compression pressure allowed by the quartz-to-metal seal), compared to a maximum of $-50$\degc{} previously.

\subsection{Chamber control and instrumentation}
\label{ss:instrumentation}

The thermodynamic state (temperature and pressure) of a bubble chamber determines the threshold energy required for bubble nucleation. For XeBC, there are two temperature regions to consider: the warm active region becomes superheated when the chamber pressure drops to the target expanded state, while the jar seal, plumbing, and the bellows that transmits pressure changes from the hydraulic fluid to the target fluid are kept cold so that liquid contacting these surfaces will not boil \cite{DBCinstrumentationpaper}. The entire chamber assembly is cooled by a cold finger running from an open-mouth dewar of liquid nitrogen to the top of the cold (inactive) region of the bubble chamber.
The warm region is then heated with a nichrome wire heater mounted on the outer wall of the copper, aluminum, and sapphire radiation shield that encloses the target volume. This shield is thermally coupled to the outer jar by an indium contact immediately below the target volume. The temperature of the warm region is controlled with feedback from a 100-$\Omega$ platinum RTD mounted on the inside of the radiation shield (T2 in Fig.~\ref{fig:xebcschematic}). Additional RTDs (T1, T3, T4) monitor temperatures throughout the warm and cold regions.

Pressure is controlled by a hydraulic system filled with Dynalene-MV, a commercial low-temperature / low-viscosity hydrocarbon blend, using a pair of hydraulic accumulators as ``reservoirs'' for the high (compressed) and low (expanded) pressures. Hydraulic pressure is transmitted to the xenon by an in-line pressure-balanced bellows. A second bellows transmits the pressure back to a warm hydraulic stub, and the pressure of all three volumes (hydraulic system output, xenon, and hydraulic stub) are monitored by absolute pressure transducers with $\sim$0.01~bar resolution (PT5, PT4, and PT6, respectively). The redundant pressure readings provide cross-calibrations to compensate for temperature-dependent readings in the cold xenon pressure transducer and for pressure differentials across the bellows due to the bellows spring constants.

Bubble chamber data is organized by ``event,'' i.e., one cycle from compressed (insensitive) $\rightarrow$ expanded (superheated, sensitive) $\rightarrow$ trigger $\rightarrow$ compressed. During a run the chamber is set to cycle automatically, repeating the cycle with a 60~s compression period between events. The camera acts as the primary bubble trigger, collecting 0.3-megapixel, 8-bit images at 100 frames per second with synchronized illumination from a ring of 950~nm LEDs mounted around the camera. The frame grabber receiving the camera images broadcasts a trigger when a significant change (typ.\ 25\% full scale) is seen in a minimum number of pixels (typ.\ 12 pixels) between consecutive images. In addition to images, data recorded for each event include pressure and temperature readings throughout the event, acoustic waveforms digitized at 2.5~MHz covering 100~ms before and after the bubble trigger, and time-stamped PMT waveforms digitized at 1~GHz for scintillation pulses observed throughout the event. The PMT digitizer is triggered by a Nuclear Instrumentation Module (NIM) discriminator. This trigger is suppressed when camera illumination is on, incurring a 10\% dead time in the scintillation channel, but is otherwise independent from the bubble chamber cycle.

%triggering recompression and packaging and saving of the event's data. The 100~frame-per-second frame rate effectively limits our visual knowledge of the bubble's nucleation time to the nearest 10~ms. In addition to the camera, RTD, and pressure information gathered, the data also includes acoustic traces (recorded by a pair of custom piezoelectric transducers), and the PMT waveforms. A muon veto paddle provides coincidence discrimination, which is monitored alongside the PMT with a NIM discriminator. 

\subsection{Low-threshold data}
\label{ss:data}

The low-threshold data analyzed here were acquired over the course of five months from June to October 2017, in three distinct xenon fills. The chamber's superheated operational temperature ranged from $-55$\degc{} to $-38$\degc{}, with target expansion pressures from 1.7--3.4~bara. Data are reported in terms of the Seitz threshold of the superheated target, calculated following \cite{ERpaper} with thermophysical properties for xenon taken from NIST REFPROP~\cite{REFPROP}. Pressure and temperature were controlled in this dataset to a precision of at least 0.04~bar and 0.1$^\circ$C, corresponding to uncorrelated uncertainties on \qseitz{} of 0.8\% and 1.2\%, respectively, across datasets. Because xenon temperature was measured only indirectly (by measuring the temperature of external elements thermally coupled to the target volume), there is an additional 1$^\circ$C global temperature uncertainty, with a corresponding 12\% global uncertainty on \qseitz{}. The uncorrelated uncertainties on \qseitz{} are small compared to statistical uncertainties in the data presented below, while the only effect of the global uncertainty on \qseitz{} is an effective relabeling of fluid states. Conclusions regarding actual nuclear recoil detection thresholds and electron recoil sensitivities are unaffected, except for the conservative constraint imposed in the analysis that there is zero sensitivity to nuclear recoils with energies below \qseitz{}. For these reasons, uncertainties on the thermal state of the chamber are neglected in the analysis.

A variety of sources were deployed, including \cf{} as a fission neutron source, \ybe{} and \bibe{} as photoneutron sources, and \y{}, \bi{}, and \co{} as $\gamma$-ray sources. Table~\ref{tab:nuisances} summarizes the activities, neutron production rates, placement, and resulting neutron fluxes in the superheated target from these sources as determined by an MCNPX-PoliMi \cite{MCNP} Monte Carlo simulation. The fission source generates neutrons over a broad spectrum with an average energy of 2.3~MeV~\cite{cf252}, and is simulated using the built-in PoliMi \cf{} fission spectrum. The photoneutron sources produce nearly monoenergetic low-energy neutrons via $^{9}$Be$(\gamma,n)^{8}$Be reactions in BeO and metallic beryllium parts surrounding the $\gamma$-ray sources. The \ybe{} source neutrons are peaked at 152~keV, with approximately 0.7\% of the neutrons at higher energy (950~keV) from the low-branching-fraction 2734-keV $\gamma$-ray~\cite{YBeCollar}. The \bibe{} source produces neutrons peaked at 91~keV. Photoneutron rates and neutron spectra are calculated using MCNPX-PoliMi, using the $^{9}$Be$(\gamma,n)^{8}$Be cross sections given in~\cite{BePhotoneutron} and correcting lab-frame neutron energies with the appropriate relativistic kinematic factor (necessary due to MCNPX's mishandling of $\gamma$ momentum~\cite{CARO2016170}).

%built-in  in python with Monte Carlo simulations of the source geometries~\cite{matt_thesis}, based on the $^{9}$Be$(\gamma,n)^{8}$Be cross-sections given in~\cite{BePhotoneutron}.

%\begin{table}
%%\begin{adjustbox}{width=\textwidth}
%    \centering
%    \begin{tabular}{c|c|c|c|c}
%        Source & Cal. Date & Cal. Act. & Half-Life & Final Act.\\
%         &  & [MBq] &  & [MBq]\\
%        \hline
%%        \hline
%%          \cf & 15 Aug. 2016 & 0.0154 & 2.647~y & 0.0114 \\
%        \hline
%          \cf & 19 Dec. 2018 & 0.00831 & 2.647~y & 0.0114 \\
%          \hline
%        \y & 25 Aug. 2017 & 27.2 & 106.6~d & 20.0\\
%        \hline
%        \bi & 15 Jan. 2015 & 0.389 & 31.55~y & 0.366\\
%        \hline
%        \co & 3 Apr. 2013 & 185 & 271.7~d & 2.73\\
%        \hline
%    \end{tabular}
%    %\end{adjustbox}
%    \caption{Neutron and $\gamma$-ray calibration sources. Cal. Act. is the source activity on Cal. Date and Final Act. is the activity on Oct.\ $\mathrm{11^{th}}$, 2017, the final day of data collection. The \y{} and \bi{} source were used both as $\gamma$-ray sources, and paired with beryllium as photoneutron sources.}
%    \label{sourcestrengthtable}
%\end{table}

\begin{table*}[!t]
    \centering
    \begin{tabular}{c|c|c|c|c|c|c}
         & Activity & Distance & \multicolumn{2}{c|}{Neutron flux at target [min$^{-1}$cm$^{-2}$]} & & \\
        Source & [MBq] / [n/s]  & [cm] & \hspace{1cm} Prior \hspace{1cm} & \hspace{1cm} Fit \hspace{1cm} & Dates in 2017 & Thresholds [keV] \\        \hline
        \hline
        \cf{} & 0.0113 / 1,320 & 84.6 & $1.18 \begin{array}{l} +0.11 \\ -0.10 \end{array}$ & 
        $1.33 \begin{array}{l} +0.08 \\ -0.08 \end{array}$ & 
        16 Jun. to 5 Aug. & $0.90, 1.19, 1.89, 2.06$ \\
        \hline
        \ybe{} & 20.0 / 560 & 35.0 & $4.40 \begin{array}{l} +0.61 \\ -0.54 \end{array}$ & 
                $4.10 \begin{array}{l} +0.50 \\ -0.32 \end{array}$ & 
                26 Sep. to 1 Oct. & $\begin{array}{l}0.91, 0.97, 1.04, 1.12, \\ 1.14, 1.33, 1.44, 1.56 \end{array}$ \\
%        & & & & & 1.14, 1.33, 1.44, 1.56 \\
        \hline
        \bibe{} & 0.366 / 1.25 & 10.4 & $0.084 \begin{array}{l} +0.029 \\ -0.022 \end{array}$  & 
                $0.083 \begin{array}{l} +0.027 \\ -0.022 \end{array}$ &  
                23 Sep. to 24 Sep. & 1.14 \\
        \hline  \hline %\multicolumn{6}{c}{ } \\
        %\cline{1-4} 
        \multicolumn{3}{c|}{Target mass prior / fit [g] } & $21.1 \begin{array}{l} +2.2 \\ -2.0 \end{array}$ &
                \multicolumn{1}{c}{$23.3 \begin{array}{l} +1.7 \\ -1.5 \end{array}$} & \multicolumn{2}{|c}{ } %\\ \cline{1-4}
    \end{tabular}
    \caption{\label{tab:nuisances}Overview of neutron sources used for nuclear recoil calibration, with pre- and post-fit values and 1-$\sigma$ uncertainties on the corresponding nuisance parameters. Activities and fluxes are referenced to October 11, 2017, the final day of data taking, and are calculated using MCNPX-PoliMi~\cite{MCNP}. The ``Distance'' column indicates the distance from the active element in the source to the center of the visible xenon target. As described in the text, the flux from each source is given a log-normal prior, as is the mass of the visible xenon target. A fourth source, $^{57}$Co (2.73 MBq), was used for gamma calibrations only.}
\end{table*}

Table~\ref{tab:nuisances} also shows uncertainties in neutron flux from each source, stemming from uncertainties both in source strength and in neutron propagation.
The \cf{} source neutron emission rate was calibrated to 6\% precision with a well-characterized $^3$He counter, and was found to be consistent with the high-energy ($>$2~MeV) gamma emission rate measured by a NaI crystal. The total uncertainty on neutron flux from the \cf{} source at the chamber is 9\% after including geometric uncertainties in neutron propagation, which were estimated by varying selected details in the simulated geometry and by comparing the MCNP-PoliMi simulation with an independently developed Geant4~\cite{Geant4} simulation. The uncertainty in the \ybe{} source neutron emission rate is 5\%, due to a combination of geometrical uncertainties in the $^{88}$Y deposition profile in the source capsule and uncertainties on the $^{9}$Be$(\gamma,n)^{8}$Be cross section, and is consistent with the neutron emission rate measured by the same $^3$He counter used to characterize the \cf{} source. Neutron propagation uncertainties are significantly higher for \ybe{} than for \cf{}, giving a total neutron flux uncertainty of 13\%. The \bibe{} neutron flux uncertainty is larger still at 30\%, dominated by uncertainty associated with additional beryllium pieces deployed with that source in an attempt to boost the neutron emission rate. Finally, there is a common-mode 10\% uncertainty on the integrated flux of neutrons through the ``active'' xenon due to uncertainty on the volume of xenon visible to the camera.

Each neutron dataset (characterized by a single source and threshold) is accompanied by a dedicated background measurement taken at the same threshold and in the same xenon fill. In the case of \cf{} this simply involves removing the source. For the \ybe{} and \bibe{} experiments, the Be and BeO components are replaced with aluminum, with the \y{} and \bi{} sources still present to preserve the same $\gamma$-ray flux incident upon the detector but without the photoneutrons. The \bi{} and \y{} $\gamma$-ray sources without beryllium are also used to measure ER sensitivity, along with the lower-energy \co{} source, at Seitz thresholds between 0.50~keV and 1.56~keV. Sources were placed 10~cm from the active xenon for all ER sensitivity measurements.

%\begin{table}
%%\begin{adjustbox}{width=\textwidth}
%    \centering
%    \begin{tabular}{c|c|c}
%        Source & Dates in 2017 & Thresholds [keV]\\
%        \hline
%        \hline
%          \cf{} & 16 Jun. to 5 Aug. & 0.90, 1.19, 1.89, 2.06 \\
%          \hline
%        \ybe{} & 26 Sep. to 1 Oct. & 0.90, 0.97, 1.04, 1.12, \\
%        & & 1.14, 1.33, 1.44, 1.56 \\
%        \hline
%        \bibe{} & 23 Sep. to 24 Sep. & 1.14 \\
%        \hline
%    \end{tabular}
%    %\end{adjustbox}
%    \caption{Neutron source deployment dates and threshold ($Q_{\text{Seitz}}$) set-points.}
%    \label{tab:thresholds}
%\end{table}

The reduced data for the subsequent analysis includes only live times and bubble counts at each threshold and source configuration (with the exception of the \cf{} data, which includes the scintillation veto described in the next section). The chamber is considered live only when the measured pressure is within 0.034~bar (0.5~psi) of the target pressure set-point (extended to $\pm$0.068~bar for the 0.90~keV \ybe{} dataset), and any live time in the the first 20~s of each event is discarded to avoid thermodynamic transients associated with expansion. Bubbles are identified through hand-scanning of images, counting bubbles that are nucleated in the visible volume and are coincident with the bubble chamber trigger. No excess rate of bubble nucleation at the wall of the chamber was observed, so no fiducial cut is taken.

\subsection{Coincident scintillation veto}
\label{sec:coin_scin}

While the scintillation signals produced by NR events can in principle be used for energy reconstruction, in practice NR photon yields are too low to be useful for this purpose. Both in XeBC and in SBC's planned dark matter and CE$\nu$NS detectors, any practical light collection configuration yields $<1$ photon detected for the $\mathcal{O}$(keV) and sub-keV nuclear recoils in these experiments' regions of interest. However, the scintillation signal still provides a useful method of background suppression, easily tagging alpha decays and transiting charged particles from cosmic-ray showers, which dominate the background bubble nucleation rate in XeBC.

The scintillation trigger is set to be sensitive to single photoelectrons in XeBC's single PMT with $>50$\% efficiency in all data where the scintillation veto is used, verified using a pulsed blue LED. Light collection efficiency is measured by exposing the chamber to 122-keV $\gamma$-rays from a \co{} source. The 122~keV photoabsorption peak is detected at $\sim$4 photoelectrons, corresponding to a total photon detection efficiency of $\sim$0.05\%. This efficiency is approximately a factor of 8 lower than reported in Ref.~\cite{XeBC}, and is thought to be due to decreased UV transparency in the new quartz vial. With this efficiency, nuclear recoils around 
%200~keV result in 1~phe collected on average, assuming the Lindhard model with biexcitonic quenching~\cite{LUX:2017bef,Mei:2007jn}. \textcolor{red}
120~keV result in one photoelectron collected on average, based on the zero-field nuclear recoil light yields given in NEST~v2.2.0~\cite{NEST}.
%{\bf Let's replace this number with one from NEST, with the appropriate NEST reference.}

Scintillation pulses associated with bubble nucleation are identified by coincidence with the onset of the acoustic signal, as in~\cite{XeBC}. The relocation of the acoustic sensors to the bottom of the chamber, however, complicates the position dependence of the observed acoustic onset, so unlike the previous analysis, no position-dependent correction is attempted. This results in a coincidence window of 450~\si{\us} rather than the 50~\si{\us} used previously. The larger window and corresponding accidental coincidence rate prevents the association of scintillation pulses with bubbles in any data where a $\gamma$-ray source is present, but bubble-coincident scintillation can still be identified in \cf{} and background data.

\begin{figure}
    \centering
    \includegraphics[trim={0.7cm 0.0cm 1.7cm 0.2cm},clip,width=\linewidth]{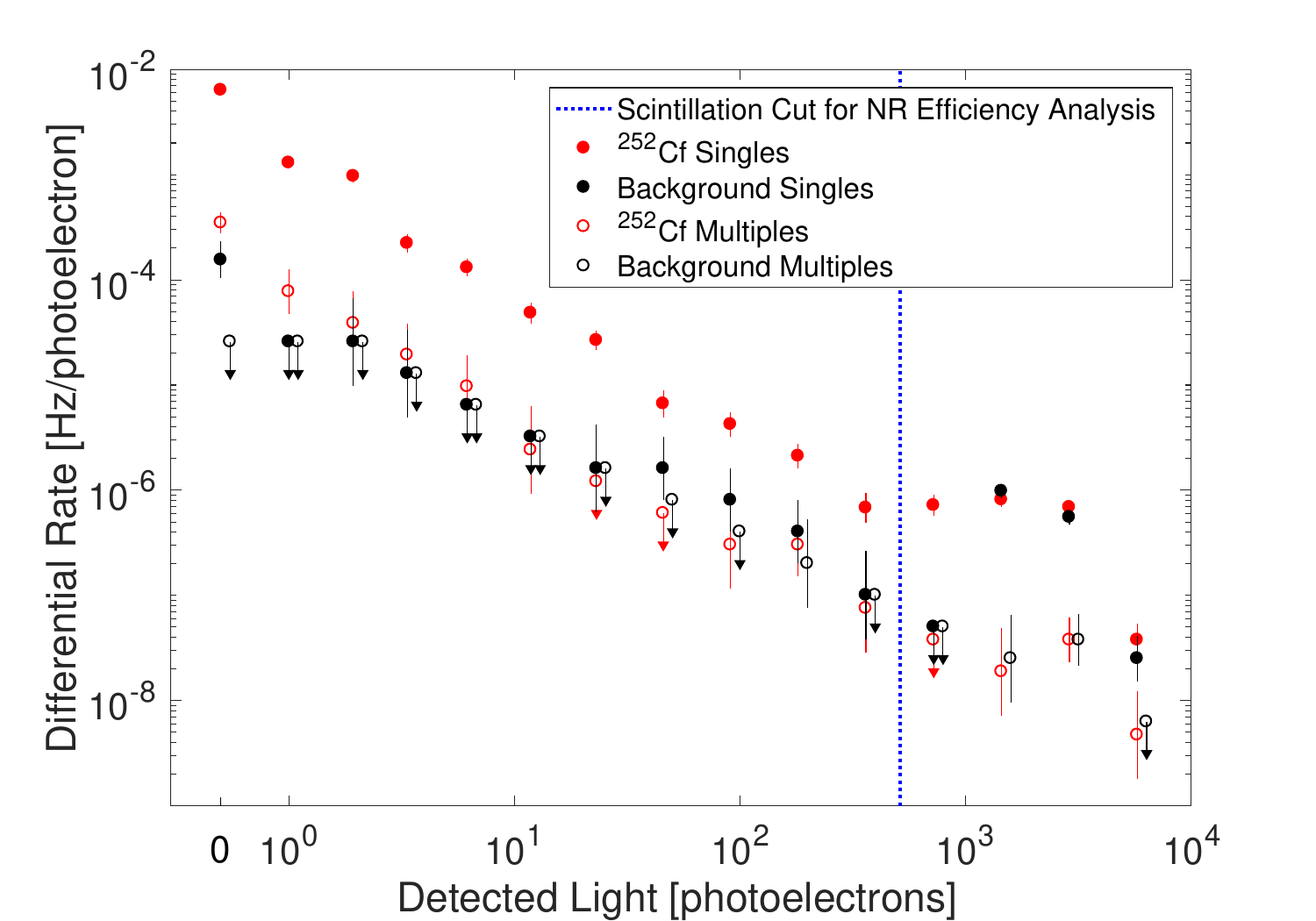}
    \caption{Scintillation spectra for \qseitz=1.19~keV \cf{} and background data. Differential rate in Hz/photoelectron is plotted for bins spanning powers of 2, with the addition of a bin for no associated light (units of Hz). Additional datasets taken at \qseitz=0.9~keV, 1.89~keV, and 2.06~keV are similar. Single-bubble events (indicating a single scatter or decay in the chamber) are shown as filled circles, while multiple-bubble events (indicating a neutron scattering more than once on its path through the chamber) are shown as unfilled circles. Markers with down-pointing arrows indicate $1\sigma$ Poisson upper limits in bins where no bubbles were observed.
    Background multiple-bubble points are shifted to avoid overlap with single-bubble points.
    %For background bins where no single- or multiple-bubble events are observed the upper limits perfectly overlap.
    A majority of bubbles in the background dataset are accompanied by large scintillation pulses, whereas the majority of the \cf{} bubbles are accompanied by no light or pulses of less than ten photoelectrons. The scintillation veto cut used in the NR efficiency analysis is shown in blue.}
    \label{cfandbgspect}
\end{figure}

Figure~\ref{cfandbgspect} shows the scintillation spectra associated with single- and multiple-bubble events in \cf{} and background data. The \cf{} source emits neutrons with energies up to 10~MeV \cite{cf252}, giving an elastic NR energy spectrum that peaks well below the 120~\kevnr{} light collection threshold. % (based on Geant4 simulations, see section \ref{sec:thresholds}), so the largest fraction of the bubble rate induced by this source is expected to be associated with no PMT pulses. The high-energy tail of the neutron energy distribution is expected to create bubbles with small scintillation pulses of 1--2 photoelectrons.
Larger scintillation pulses of tens of photoelectrons accompany inelastic scatters, where the scintillation is dominated by ERs produced through nuclear relaxation. Background single-bubble events from $\alpha$-decays and cosmic rays give much stronger scintillation, with $\sim$85\% of background events coincident with scintillation pulses of $>10^3$~photoelectrons. A cut at 512~photoelectrons reduces the single-bubble background rate from $\sim$0.2 bubbles/minute to $\sim$0.02 bubbles/minute without impacting nuclear recoils from \cf{}.

%Measured rates are shown in Fig.~\ref{cfandbgspect}; the spectrum for data with the \cf{} source does peak at zero, with significant rates above background in the bins below 10~phe. The background scintillation spectrum for this experiment is also shown in Fig.~\ref{cfandbgspect}. The background spectrum is dominated by high-phe events ($\sim$85\% are above 1000~phe), which are also present at a similar rate in the \cf{} data. Based on these results, events above 512.5 phe (corresponding to a bin boundary in Fig.~\ref{cfandbgspect}) are cut, to remove high-phe background events from the \cf{} data.

%The PMT pulse-bubble coincidence finding algorithm is vetted in part by observing the $z$-dependence of the difference between the acoustic signal arrival time to the piezo and the light arrival time to the PMT. The sound travel time between the nucleation site and the piezo is measurable against the essentially instantaneous arrival of the light to the PMT. Sound speeds around 500~m/s are observed, within uncertainties with literature values of the sound speed in liquid xenon \cite{xenonsoundspeed}.

\section{Results}
\label{S:results}
\subsection{Nuclear recoil nucleation efficiency}
\label{sec:thresholds}

\begin{figure}
    \centering
    \includegraphics[width=0.96\columnwidth]{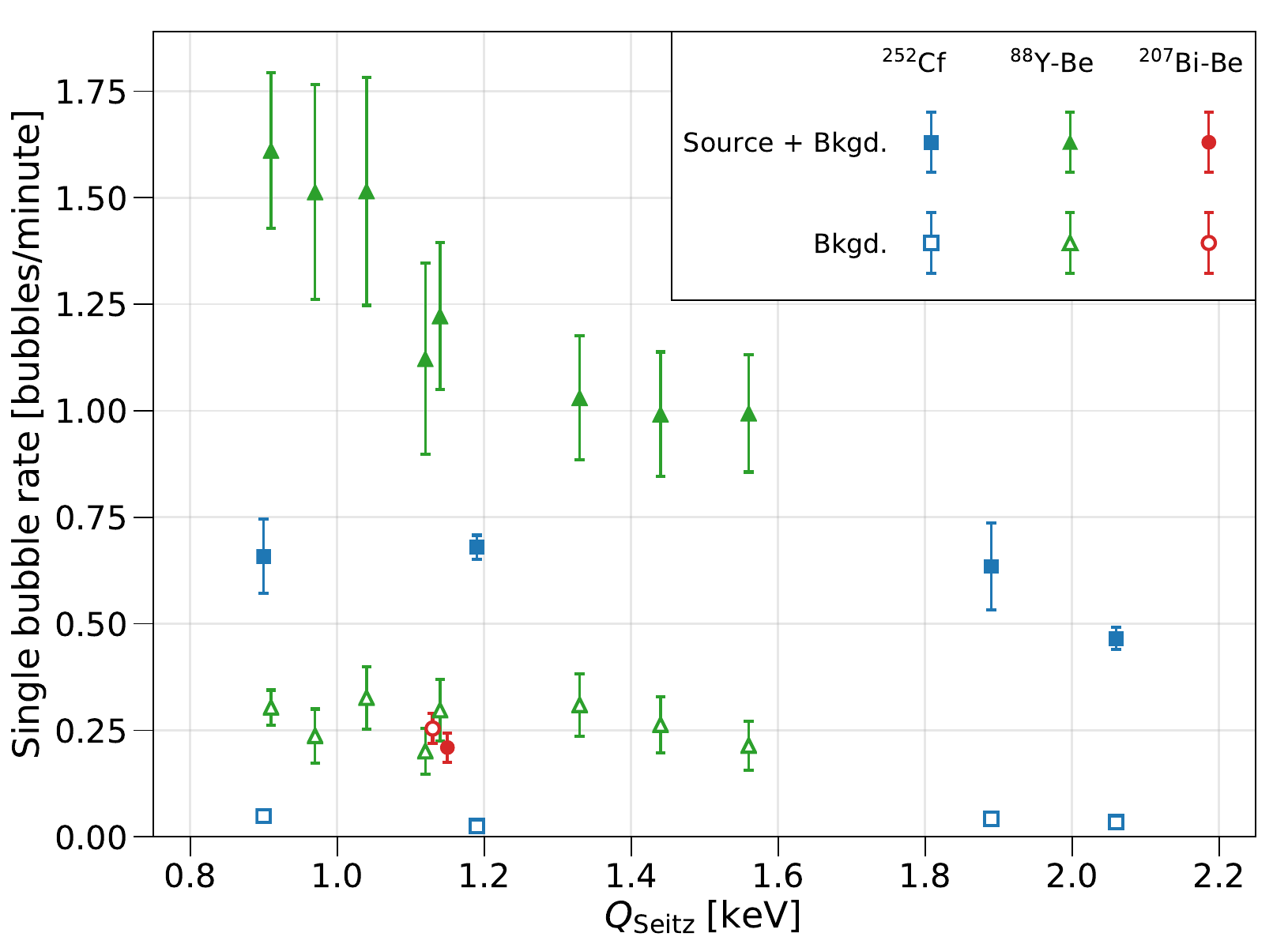}
    \caption{Rate of single-bubble counts (after scintillation veto in the case of \cf{} and associated background data) as a function of thermodynamic threshold \qseitz{}, for all nuclear recoil calibration experiments. The error bars represent statistical uncertainty. Empty markers indicate background-only data corresponding to each experiment. Note that the overlapping \bibe{} data points are displaced in \qseitz{} by $\pm 0.01\,\mathrm{keV}$ (from $1.14$~keV) for clarity.}
    \label{fig:rate_vs_thresh}
\end{figure}

The measured neutron source and background single-bubble rates are shown as a function of \qseitz{} in Fig.~\ref{fig:rate_vs_thresh}. The XeBC demonstrates an increased rate of bubbles over the background when exposed to \cf{} and the \ybe{} photoneutron source, but no excess rate from the much weaker \bibe{} photoneutron source. The lack of evidence for nucleation by the weak \bibe{} source is to be expected -- the original intent was to apply the scintillation veto to \bibe{} data to suppress the observed background, but as described above, the increased coincidence window precluded the use of the veto with a $\gamma$-ray source present. Bubble rates observed in background data are roughly constant over the range of \qseitz{} explored here, while the rates of bubbles observed with \ybe{} and \cf{} decrease as \qseitz{} increases and the sensitivity to low-energy recoils diminishes. The maximum xenon recoil energy for NRs induced by 152~keV \ybe{} neutrons is 4.6~keV. This gives an immediate upper bound on the actual bubble-nucleation threshold; because there is a rate above background, there must be some sensitivity to recoils below 4.6~keV.

Figure~\ref{fig:rate_vs_thresh} illustrates nucleation rates versus the Seitz threshold for fixed nuclear recoil spectra, but physics results from dark matter searches or CE$\nu$NS measurements depend on knowing bubble nucleation efficiency as a function of recoil energy at a fixed thermodynamic state. Some past efforts at bubble chamber calibration have performed this conversion by assuming specific functional forms for nucleation efficiency that link dependence on recoil energy to dependence on the Seitz threshold~\cite{COUPP4,Tardif}, but the physical basis for such assumptions is unclear. While ongoing theoretical efforts attempt to better constrain nucleation efficiency models using a combination of recoil cascade simulations and molecular dynamics evolutions~\cite{lammps_paper,bubbleMD}, recent experimental efforts have shifted toward purely data-driven constraints on nucleation efficiency that gain their constraining power from measurements of multiple recoil spectra, rather than measurements taken at multiple thermodynamic states. The analysis presented here takes the approach developed in \cite{PICONR}, making the same minimal set of assumptions: thermodynamic states with the same \qseitz{} are assumed to give the same nuclear recoil response (an assumption supported by data in \cite{PICONR} for the range of pressures considered in this work); efficiency is assumed to increase as recoil energy increases or \qseitz{} decreases; and efficiency is assumed to be zero when recoil energy is less than \qseitz{}.

To obtain a complete estimate of the NR nucleation efficiency as a function of recoil energy $E_r$ and \qseitz{}, a global analysis using all neutron calibration data gathered with the chamber is undertaken, closely following the analysis performed in Ref.~\cite{PICONR}. The data is binned by bubble count in each event (``multiplicity''), \qseitz{}, and calibration source. The background data is binned in the same way, with a 1-to-1 map between source and background datasets. For consistency, the background and source data are required to be taken during the same xenon fill.

To extract a nucleation efficiency measurement, the expected energy spectrum and multiplicity of neutron interactions in the detector is required; this information is extracted from the detailed PoliMi output of the MCNPX-PoliMi simulations described in Sec.~\ref{ss:data}. In the limit where neutrons scatter at most once in the detector, the expected experimentally observed bubble rate $R_{\rm{obs}}$ can be calculated for a given bubble nucleation efficiency model $\epsilon(Q_{\text{Seitz}},E_r)$ as
\begin{equation}
%\label{eq:bubble_rate}
    R_{\rm{obs}}  = \int_0^\infty R_{\rm{MC}}(E_r)\cdot\epsilon(Q_{\text{Seitz}},E_r)\,dE_r,
\end{equation}
where $R_{\rm{MC}}(E_r)$ is the simulated energy spectrum. To include multi-scatter events, a probabilistic approach is used, turning each simulated recoil into a bubble with probability $\epsilon(Q_{\text{Seitz}},E_r)$. For this analysis, simulated and real data are divided into two multiplicity bins, counting single- and multi- (2+) bubble events.

As no \emph{a priori} functional form is known for the efficiency function $\epsilon(Q_{\text{Seitz}},E_r)$, a generic model of piecewise linear functions is used. Specifically, the efficiency function is modeled as a triangular mesh, with triangle vertices at fixed efficiencies (0, 0.2, 0.5, 0.8, and 1) and fixed \qseitz{} (0.90~keV and 2.06~keV). Each rectangle in this grid is divided into two triangles by the diagonal from (low-$\epsilon$, low-\qseitz{}) to (high-$\epsilon$, high-\qseitz{}). The position of each vertex in $E_r$ is allowed to vary -- giving 10 free parameters -- but constrained such that the resulting efficiency function is monotonic in both $E_r$ and \qseitz{}, and gives $\epsilon=0$ when $E_r<Q_\mathrm{Seitz}$. The vertex positions in efficiency are chosen to match Ref.~\cite{PICONR}, and the vertex positions in \qseitz{} are chosen to match the highest/lowest \qseitz{} in the data. Models with additional equally spaced vertices in \qseitz{} are also considered.
Choosing the optimal number of threshold (\qseitz{}) set-points is a balance between having a sufficiently flexible model while not overfitting. To determine the ideal solution, the data is fit with models defined with up to five threshold set-points. The Akaike information criterion \cite{AIC_modelSelection} is used to compare models with two or more threshold set-points, with two found to be the preferred number.%; $Q_{\text{Seitz}} = 0.9\,\mathrm{keV}$, $1.48\,\mathrm{keV}$, and $2.06\,\mathrm{keV}$.

In addition to the ten parameters of interest defining the nucleation efficiency model, nuisance parameters are included to reflect the neutron flux uncertainties from each source (9\%, 13\%, and 30\% for the \cf{}, \ybe{}, and \bibe{} sources, respectively) as well as a 10\% common-mode uncertainty reflecting uncertainty on the mass of the visible xenon target. Each of these four nuisance parameters is implemented as a multiplier on the expected rates in the corresponding datasets.

Constraints on the efficiency curve and nuisance parameters are found using a binned maximum likelihood with background subtraction. This likelihood is given (up to a multiplicative constant) by
\begin{equation}
\begin{aligned}
\mathcal{L}&\big(\vec{\theta},\vec{s}\, \big| \, \{n_{i,j}\},\{m_{i,j}\},\{r_i\}\big) = \prod_ke^{-\frac{s_k^2}{2}} \,\,\times \\
&     \prod_{i,j}
    e^{-\mu_{i,j}}\mu_{i,j}^{m_{i,j}}
    \cdot 
    e^{-(\nu_{i,j}+r_i\cdot\mu_{i,j})}(\nu_{i,j}+r_i\cdot\mu_{i,j})^{n_{i,j}},
\end{aligned}
\end{equation}
%\textcolor{red}{\bf alternate formula, choose one} \\
%(normalized so that $\mathcal{L}\le1$) by
%\begin{equation}
%\begin{aligned}
%\mathcal{L}\big(\vec{\theta},\vec{s}\, &\big| \, \{n_{i,j}\},\{m_{i,j}\},\{r_i\}\big) = \prod_ke^{-\frac{s_k^2}{2}} \,\,\times \\
%&     \prod_{i,j}
%    e^{m_{i,j}-\mu_{i,j}}\left(\frac{\mu_{i,j}}{m_{i,j}}\right)^{m_{i,j}}
%    \,\,\times \\ & \prod_{i,j} 
%    e^{n_{i,j}-(\nu_{i,j}+r_i\cdot\mu_{i,j})}\left(\frac{\nu_{i,j}+r_i\cdot\mu_{i,j}}{n_{i,j}}\right)^{n_{i,j}},
%\end{aligned}
%\end{equation}
where $n_{i,j}$ ($m_{i,j}$) are the number of events observed with (without) the neutron source present in configuration $i$, falling in bubble-multiplicity bin $j$. The $r_i$ give the relative live time of source-on to source-off data in each configuration. The floating parameters $\vec{\theta}$ and $\vec{s}$ indicate the efficiency curve and nuisance parameters, respectively, the latter each given a Gaussian prior with mean of zero and standard deviation one (giving a log-normal prior on the physical uncertainties they represent). The $\nu_{i,j}(\vec{\theta},\vec{s})$ are the expected (i.e., simulated) number of bubble events generated by the source in configuration $i$ in bubble-multiplicity bin $j$, given a particular efficiency curve and nuisance parameter set. Finally, the $\mu_{i,j}$ are implicit nuisance parameters for the background rate in each configuration and multiplicity bin. Given a hypothesis $\nu_{i,j}$, the optimum $\mu_{i,j}$ is found analytically by solving $\frac{\partial\mathcal{L}}{\partial\mu_{i,j}}=0$. This leads to the non-negative, real solution
\begin{equation}
    \mu = \frac12\left(
    \frac{n+m}{1+r}-\frac{\nu}{r}\right)+
    \sqrt{\frac14\left(\frac{n+m}{1+r}-\frac{\nu}{r}\right)^2+\frac{\nu}{r}\frac{m}{1+r}},
    %\right)
\end{equation}
where subscripts have been left off for clarity.
%\begin{equation}
%    \mu^2 + \left(\frac{\nu}{r}-\frac{n+m}{1+r}\right)\mu - \frac{\nu}{r}\frac{m}{1+r}=0,
%\end{equation}
%which always has one positive real solution.
%When the background is well constrained (\emph{i.e.}, source-off livetime much greater than source-on livetime, or $r\rightarrow0$) this reduces to $\mu=m$, as expected.
Finally, the likelihood can be interpreted as a chi-square statistic by taking
%\begin{equation}
    $\chi^2 = -2\ln{\mathcal{L}}$, % + C,
%\end{equation}
where $\mathcal{L}$ is normalized such that a zero residual gives zero contribution to $\chi^2$.

\begin{figure}
    \centering
    \includegraphics[width=\columnwidth]{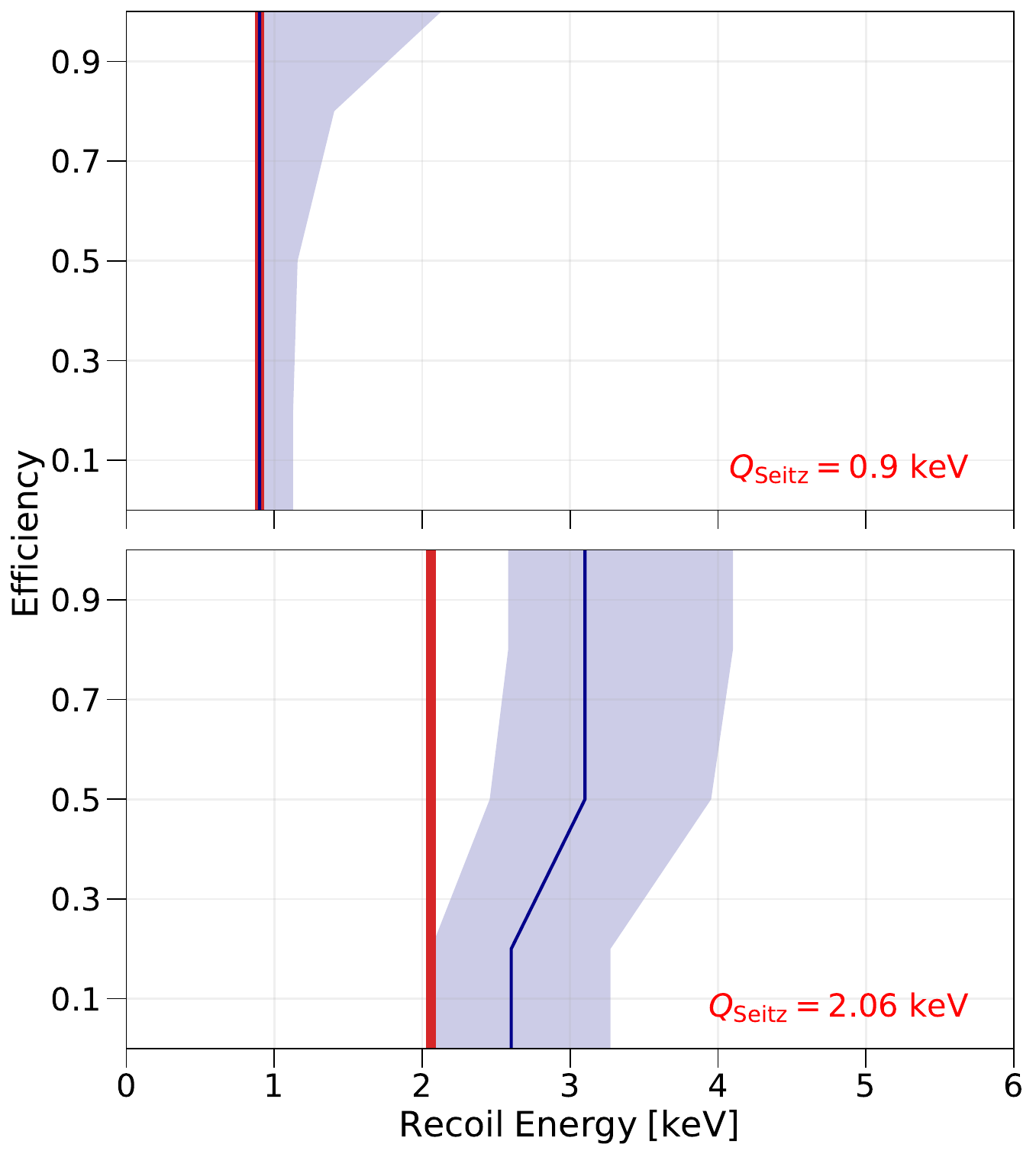}
    \caption{Best-fit nucleation efficiency model (dark blue curves) and $1\,\sigma$ uncertainty bands at the model's two threshold set-points. The corresponding $Q_{\mathrm{Seitz}}$ thresholds are shown in red. The model is constrained to give zero efficiency for recoils with energy less than $Q_{\mathrm{Seitz}}$.}
    \label{fig:nucleation_fit}
\end{figure}

Fitting the (10+4)-dimension model to the data is done with a custom Markov chain Monte Carlo (MCMC) \cite{emcee}, using the algorithm described in Refs.~\cite{jin_thesis,PICONR, ddurnford_thesis} to determine both the best fit and the boundary of the 1-$\sigma$ region around the best fit. The nucleation efficiency model obtained from this analysis is shown in Fig.~\ref{fig:nucleation_fit}.% for the two thermodynamic set-point fence posts. 

As found in previous measurements, the threshold for nucleation may deviate from $Q_{\text{Seitz}}$. For the lower thermodynamic set-point, the best-fit threshold curve closely coincides with $Q_{\text{Seitz}}$, the lower bound imposed on the model. The higher thermodynamic set-point curve is offset from $Q_{\text{Seitz}}$ by approximately 1~keV. A key result of this fit is that it indicates sensitivity to nuclear recoils of $1\,\mathrm{keV}$. Additionally, recent theoretical predictions of nucleation thresholds in various fluids using molecular dynamics simulations compare favorably with the present results \cite{lammps_paper}. 

\begin{figure}
    \centering
    \includegraphics[width=\columnwidth]{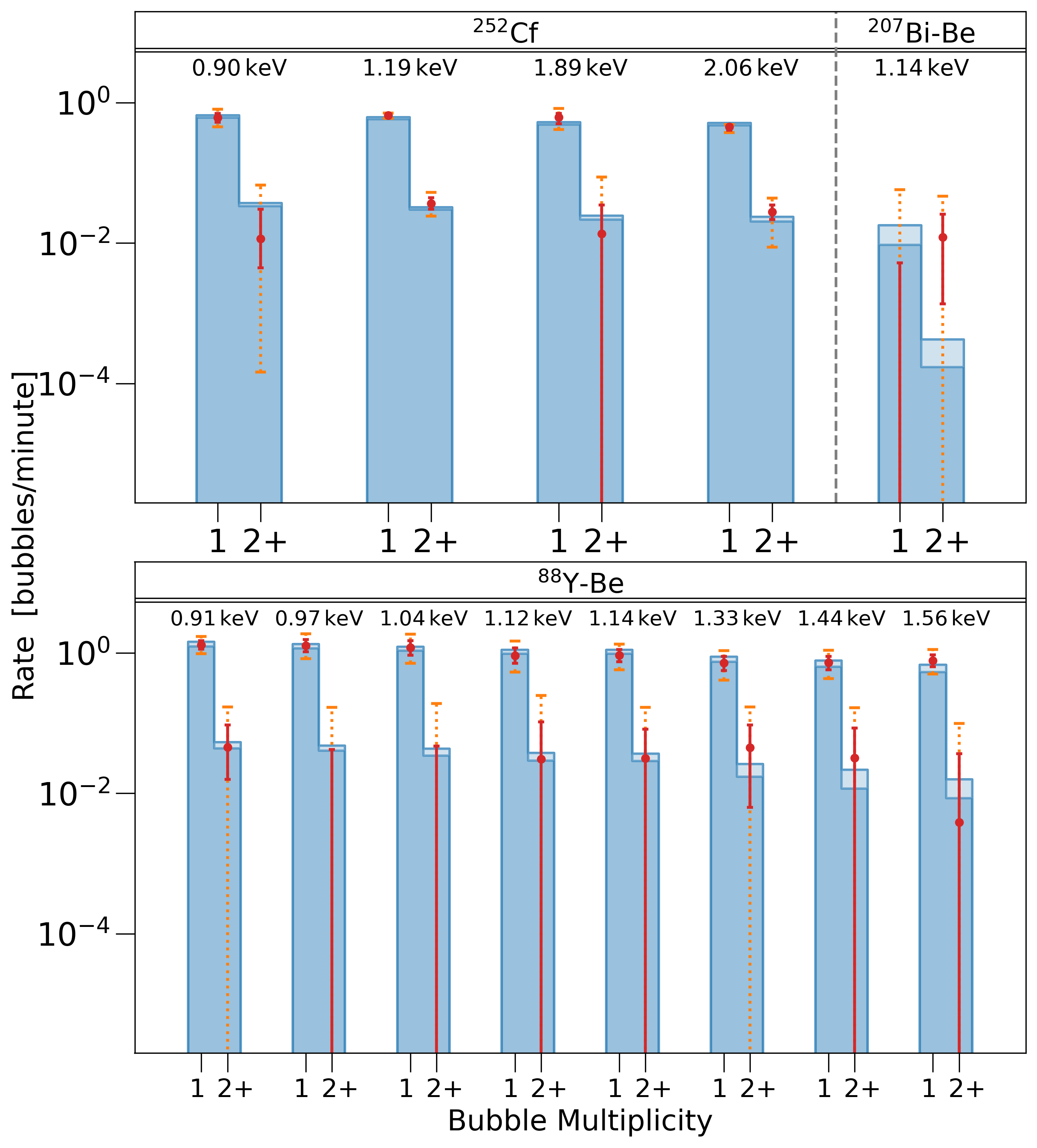}
    \caption{Background-subtracted bubble rates of the neutron calibration data (red points) separated by calibration source, $Q_{\mathrm{Seitz}}$, and bubble multiplicity. The $\pm 1\,\sigma$ (red) and $\pm 2\,\sigma$ (orange dashed) error bars of the data points include the statistical uncertainty of the signal and background. The blue histogram represents the best-fit model to the data, with the half-shaded regions showing the $\pm\,1\,\sigma$ error windows of the fit.}
    \label{fig:nucleation_gof}
\end{figure}

A direct comparison of the best-fit model and each data point is given in Fig.~\ref{fig:nucleation_gof}. As expected, the \bibe{} data proves to be of little utility, with simulated count rates that are much smaller than measured backgrounds. Multi-bubble events in \ybe{} data are also consistent with background at 2-$\sigma$ or closer in all cases, meaning that constraints on the efficiency function are driven primarily by \ybe{} single-bubble rates and \cf{} single- and multi-bubble rates. Postfit constraints on nuisance parameters, given in Table~\ref{tab:nuisances}, also show reasonable agreement with nominal values (all within 1.5 standard deviations).

\begin{figure}
    \centering
    \includegraphics[width=0.9\columnwidth]{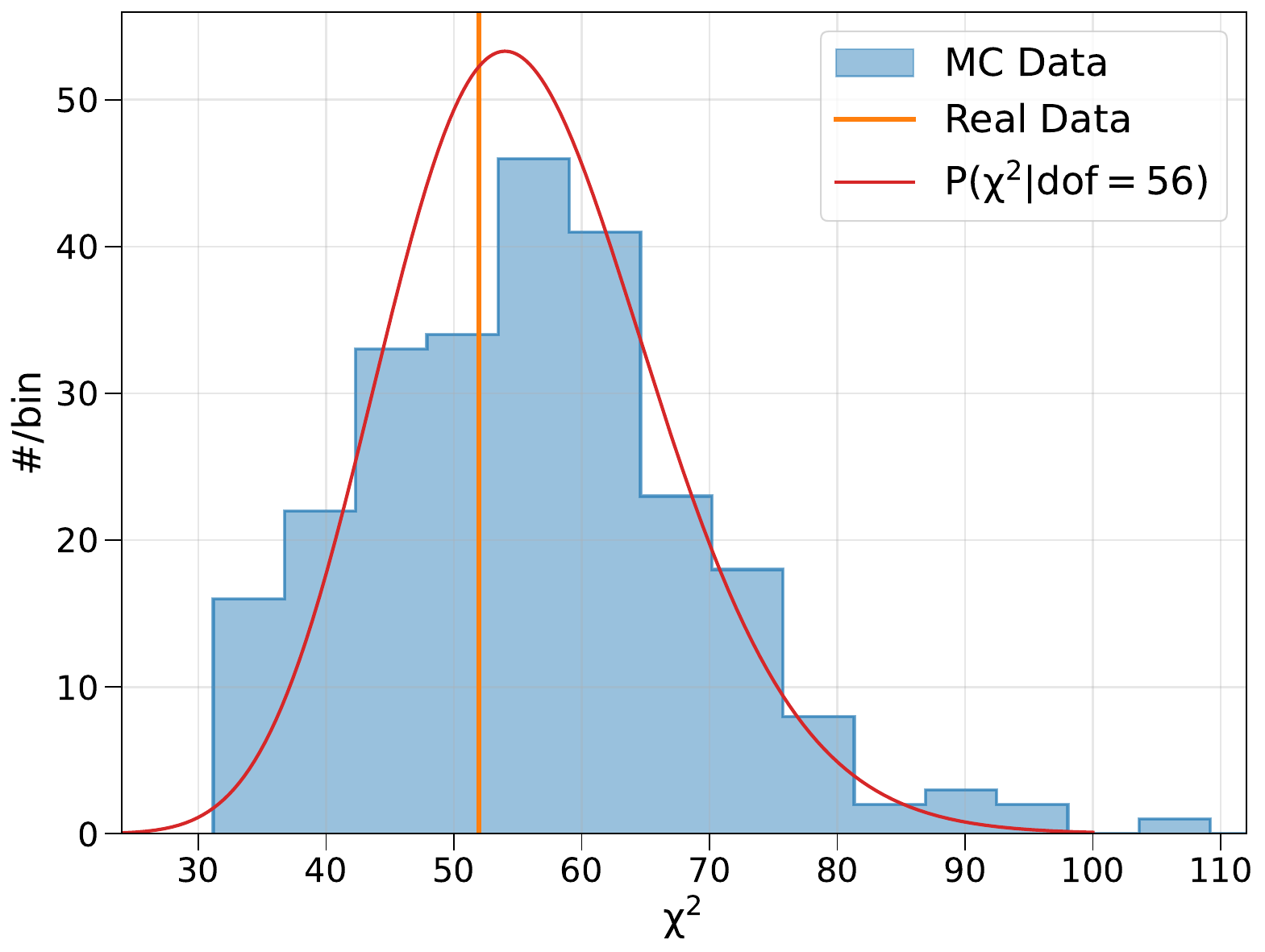}
    \caption{The distribution of $\chi^2$ for 250 simulated datasets (blue histogram) fit with a $\chi^2$ distribution with 56 degrees of freedom (red curve), compared to the observed value of $\chi^2 = 51.97$ for the fit result of the real data (orange line).}
    \label{fig:xenr_chi2}
\end{figure}

To assess the quality of the fit result, a Monte Carlo study is carried as in Ref.~\cite{PICONR}. Simulated datasets are generated using the best-fit model presented above, with random background and signal+background counts drawn for all experiments. There are 250 such datasets generated, and then fit using the same procedure used for the real data. The resulting distribution of minimum $\chi^2$ for each fit is shown in Fig.\ \ref{fig:xenr_chi2}, which is best fit by a $\chi^2$ distribution with 56 degrees of freedom. The distribution of $\chi^2$ of the MC datasets is then used to calculate a $p$-value of $0.62$ for the best-fit model of the real data. This suggests that the generic model used in this analysis does adequately represent the data, and that the fit obtained is converged. Additional Monte Carlo studies to validate the model used in this work -- in particular to demonstrate that this analysis paradigm produces unbiased results -- were carried out in Ref.~\cite{ddurnford_thesis}.

\subsection{Gamma-ray-induced bubble nucleation}
\label{gammas}

Raw bubble rates in the presence of the $\gamma$-ray sources are calculated in the same way as the rates for the nuclear recoil and background data described previously. Background-subtracted rates for $\gamma$-ray sources are calculated only for datasets that are accompanied by a background measurement under the same conditions, and only if both the $\gamma$-ray source and background datasets have greater than ten bubbles. These requirements are met only for data at thresholds $\ge$0.90~keV, due to challenges acquiring significant live time at the fixed threshold in this device at higher degrees of superheat. Data at thresholds from 0.50--0.90~keV are still used without background subtraction to give upper limits on $\gamma$-induced bubble nucleation rates. %Statistical uncertainties are added in quadrature.

\begin{figure}
    \centering
    \includegraphics[width=\columnwidth]{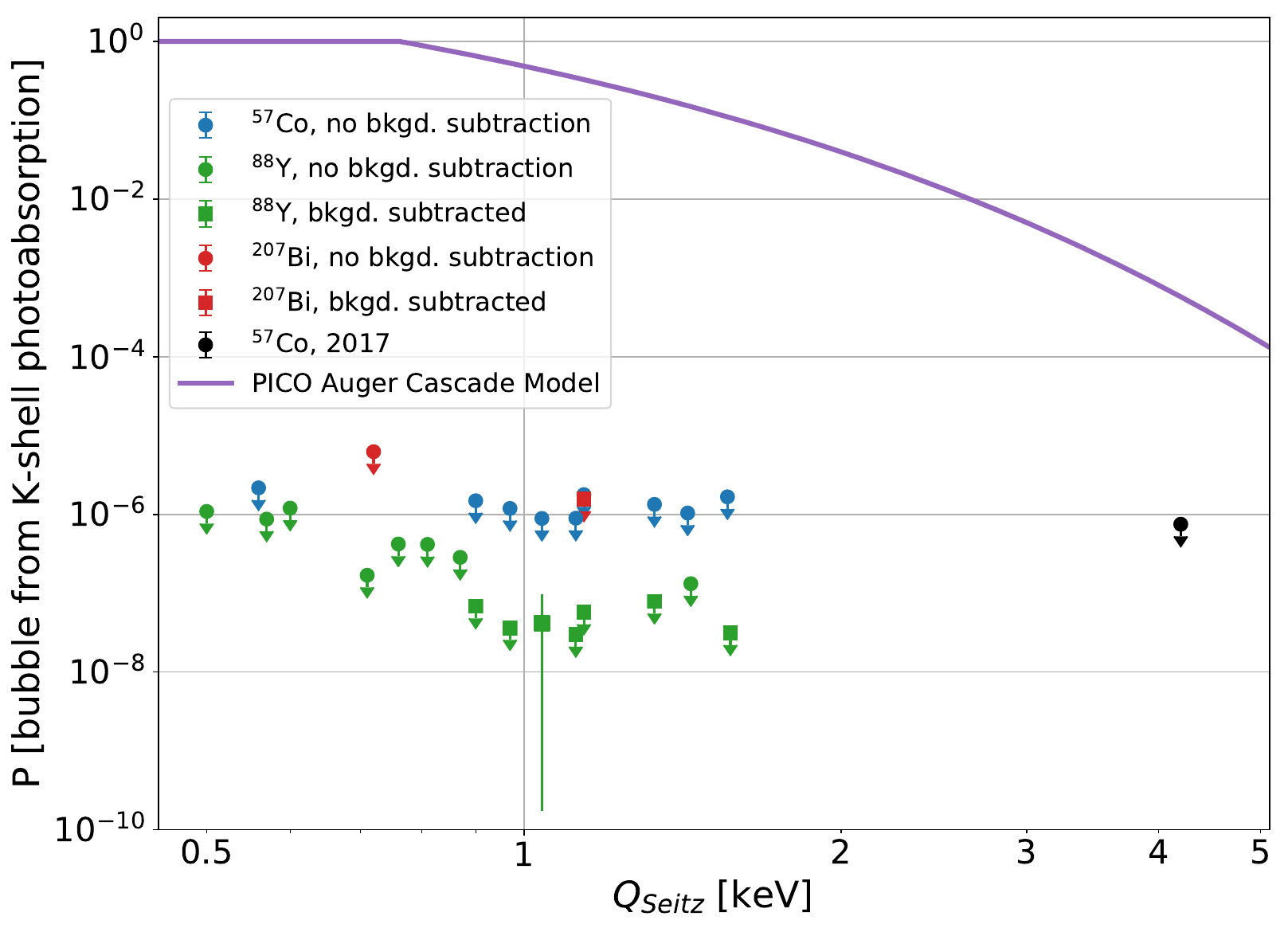}
    \caption{90\% confidence level upper limits on the probability of bubble nucleation per K-shell $\gamma$-ray photoabsorption obtained with \co{}, \y{}, and \bi{} data (blue, green, and red markers, respectively). Background is subtracted only for datasets accompanied by a background measurement. These are compared to the PICO model for nucleation by xenon contamination in C$_3$F$_8$ at 1.7~bara (Ref.~\cite{DBCAuger}, purple curve), and the previous \co{} XeBC limit presented in Ref.~\cite{XeBC} (black marker). The \y{} data of this work produce the strongest upper limits, due to the high activity of the source.}
    \label{gammarejectionplot}
\end{figure}

No significant excess over the background rate is observed. To set upper limits on the $\gamma$-ray-induced nucleation probability, the rates of photoabsorptions in the chamber for each source are calculated with Geant4 simulations; for a heavy element such as xenon, K-shell photoabsorption is expected to be the dominant mechanism for $\gamma$-ray-induced nucleation. 
%The 90\% confidence level upper limits on $\gamma$-ray-induced bubble nucleation probabilities $P_{90\%}$ are given in Table~\ref{gammarejectiontable}. 
Figure~\ref{gammarejectionplot} shows the corresponding 90\% confidence level upper limits on the probability of bubble nucleation per K-shell photoabsorption. When a background-subtracted rate is available, the upper limit is set using the signal-greater-than-zero Gaussian table of Feldman and Cousins \cite{FCpaper}. Otherwise, Poisson upper limits are presented without background subtraction.

The strongest upper limits -- of $\mathcal{O}(10^{-7})$ nucleation probability -- are produced by \y{} data, due to the high activity of the source. At the lowest achievable Seitz threshold with this setup, 0.50~keV, an upper limit of $1.1\times10^{-6}$ is set using \y{} without background subtraction. Sensitivities for the background-subtracted measurements, calculated according to the Feldman-Cousins procedure \cite{FCpaper} by simulating 1000 trials per experiment and measuring the average upper limit attained in the simulated trials, are consistent with the observed limits.

These results may be compared to the PICO Auger-cascade bubble nucleation model~\cite{DBCAuger} for xenon-contaminated C$_3$F$_8$ at an operating pressure of 1.7~bara, which provides a prediction for the nucleation rate per xenon photoabsorption as a function of \qseitz{}, shown in Fig.~\ref{gammarejectionplot}. The $\gamma$-ray rejection in xenon is found to be at least 6 orders of magnitude stronger than the projected value from xenon-contaminated C$_3$F$_8$ for thresholds around 1~keV. 

\section{Conclusion and Discussion}

The above results show that the relationship between \qseitz{} and the nuclear recoil bubble nucleation threshold, previously measured in C$_3$F$_8$ at \qseitz{} of a few keV, also holds in xenon at \qseitz{} down to at least 0.9~keV. This is significant both as it confirms the observation in \cite{XeBC} that nucleation by nuclear recoils is \emph{not} suppressed in pure xenon, and as it represents the first calibrated test of the Seitz model against nuclear recoil thresholds for $Q_\mathrm{Seitz}<1$~keV in any fluid. This measurement also adds support for the use of molecular dynamics (MD) simulations of bubble nucleation in Lennard-Jones fluids as an increasingly useful predictor of nucleation efficiency for nuclear recoils~\cite{lammps_paper,bubbleMD}, confirming the $\mathcal{O}(1)$ relationship between \qseitz{} and the nuclear recoil energy threshold in the regime where MD simulations are expected to be most reliable (i.e., low threshold in a noble liquid).
%This measurement also bridges the gap between PICO's C$_3$F$_8$ calibrations \cite{PICONR} and molecular dynamics simulations of bubble nucleation in Leanord-Jones fluids at \qseitz{} of $\mathcal{O}(100)$~eV, which show a similar $\mathcal{O}(1)$ relationship between \qseitz{} and the nuclear recoil energy threshold for bubble nucleation~\cite{lammps_paper,bubbleMD}. 

At the same time, electronic recoils continue \emph{not} to nucleate bubbles in xenon at \qseitz{} down to at least 0.5~keV, in agreement with past literature on liquid xenon and in extreme tension with electronic recoil bubble nucleation in C$_3$F$_8$. The authors suggest an interpretation of this phenomenon similar to that found in \cite{bolozdynya2010emission} -- namely that bubble nucleation by $\delta$-electrons and Auger cascades requires the presence of molecular degrees of freedom, which provide a channel for electrons to directly heat the fluid, e.g.\ by exciting molecular vibrational states. In the absence of these degrees of freedom, electrons can lose energy only through ionization and radiation, including through the scintillation emission noble liquids are most known for. A key open question is whether this insensitivity to electron recoils persists down to the homogeneous nucleation limit, the point at which random fluctuations in density begin to nucleate bubbles in the superheated liquid at a measurable rate. This limit is reached at Seitz thresholds of $\sim$75~eV and $\sim$40~eV in superheated xenon and argon, respectively~\cite{HomogeneousNucleation}.

These results, combined with the demonstrated use of coincident scintillation to reject high-energy bubble-nucleating events, indicate a unique potential for quasi-background-free searches for sub-keV nuclear recoils, such as a GeV-scale dark matter search or reactor CE$\nu$NS measurement as described in \cite{sbc_snowmass,universe9080346}. This is particularly true if electronic recoil insensitivity persists at still greater superheat and if these results extend to liquid argon, which at $\sim\!\frac13$ the mass of xenon is better suited to a low-momentum nuclear recoil search.

\section{Acknowledgments}
We are grateful to the Northwestern University department of Physics and Astronomy and to a number of students including Theo Baker, Jon Chen, Trent Cwiok, Allison Grimsted, Jared Gupta, Miaotianzi Jin, Jacob McLaughlin, Ricky Puig, Dylan Temples, and David Velasco for their support in running this detector. This work was supported by DOE Office of Science Grants No.\ DE-SC0015910, No.\ E-SC0017815, No.\ DE-SC0024254, and Do.\ DE-SC0011702, National Science Foundation Grants No.\ DMR-1936432, No.\ PHY-2310112, and No.\ PHY-2411655, Project No.\ CONACYT CB-2017-2018/A1-S-8960, DGAPA UNAM Grant No.\ PAPIIT IN105923, and Fundaci\'on Marcos Moshinsky. We acknowledge support from the Canada First Research Excellence Fund through the Arthur B. McDonald Canadian Astroparticle Physics Research Institute and the Natural Sciences and Engineering Research Council of Canada (NSERC). 

We also thank the Digital Research Alliance of Canada \cite{computecanada} and the Centre for Advanced Computing, ACENET, Calcul Qu\'ebec, Compute Ontario, and WestGrid for computational support. The work of M.~Bressler was supported by the Department of Energy Office of Science Graduate Instrumentation Research Award. The work of D.~Durnford and A.~de~St.~Croix was additionally supported by the NSERC Canada Graduate Scholarships -- Doctoral program.

\bibstyle{unsrt}
\bibliography{main.bib}

\end{document}